\documentclass{aa}

\usepackage{graphicx}
\usepackage{txfonts}
\usepackage{epsfig}

\begin{document}
\title{Discovery of new Milky Way star cluster candidates in the 
2MASS point source catalog III. Follow-up observations of cluster 
candidates in the Galactic Center region.
\thanks {Based on observations collected with the 6.5m
Magellan Baade telescope, Las Campanas Observatory.}
}
\subtitle{}

\author{J.~Borissova\inst{1,2}
\and
V.D.~Ivanov\inst{2}
\and 
D.~Minniti\inst{1}
\and
D.~Geisler\inst{3}
\and 
A.W.~Stephens\inst{4}
}

\offprints{J.~Borissova}

    \institute{
Pontificia Universidad Cat\'{o}lica de Chile, Facultad de F\'{\i}sica, 
Departamento de Astronom\'{\i}a y Astrof\'{\i}sica,
Av. Vicu\~{n}a Mackenna 4860, 782-0436 Macul, Santiago, Chile\\
\email{dante@astro.puc.cl}
      \and
        European Southern Observatory, Karl-Schwarzschild-Str. 2,
D-85748 Garching bei Mnchen, Germany\\
        \email{jborisso@eso.org, vivanov@eso.org}
      \and
        Grupo de Astronom\'{\i}a, Departamento de F\'{\i}sica, 
        Universidad de Concepcion, Casilla 160-C, Concepcion, Chile\\
        \email{doug@kukita.cfm.udec.cl}
      \and
        Princeton University Observatory, Peyton Hall, Ivy Lane, 
        Princeton, NJ 08544-1001\\
        \email{stephens@astro.princeton.edu}
        }

\date{Received .. ... 2004; accepted .. ... 2004}

\authorrunning{Borissova et al.}
\titlerunning{Follow up observations }

\abstract{
This paper is part of a project to search the inner Milky Way for 
hidden massive clusters and to address
the question of whether our Galaxy still forms clusters similar to the
progenitors of the present-day globular clusters.

We report high angular resolution deep near-infrared imaging of 
21 cluster candidates selected from the catalogues of Bica et al. 
(\cite{bic03a}) and Dutra et al. (\cite{dut03}) in a region around the 
Galactic Center. These catalogues were
created from visual inspection of the 2MASS images. Seven objects 
appear to be genuine clusters, and for these objects we present estimates of
extinction, distance and in some cases age and mass. 

Our estimated masses range from 1200 to 5500 $M_{\odot}$.  These
clusters are thus significantly smaller than any Galactic globular cluster, and
indicate that the formation of massive young clusters such as Arches and
Quintuplet is not common in the present-day Milky Way.

The remaining 14 objects are either not clusters or cannot be 
classified based on our data.


}

\maketitle

\section{Introduction}

Stellar clusters have long been recognized as important laboratories 
for astrophysical research. They provide homogeneous and well-defined 
samples of equidistant, coeval and chemically homogeneous stars. This has made 
them important tools for studies of star formation and evolution in the local 
Universe. 

The recent development of infrared instrumentation has led to the 
discovery of a large number of heavily obscured star clusters in the
Milky Way. Imaging surveys such as the Two Micron All Sky Survey 
(2MASS, Skrutskie et al. \cite{scr97}) and the Deep Near Infrared 
Southern Sky Survey (DENIS, Epchtein et al. \cite{epc97}) offer 
material for cluster identification. It appears that such hidden 
clusters are surprisingly numerous. For example, 
Hurt et al. (\cite{hur00}) found two new globular clusters: 
2MASS\,GC01 and 2MASS\,GC02 and very recently Kobulnicky et al. (\cite{kob04}) 
identified on the Spitzer Space Telescope images from the Galactic Legacy Infrared 
Mid-Plane Survey Extraordinaire (GLIMPSE) a low-latitude rich star cluster named
"GLIMPSE-C01". Dutra \& Bica (\cite{dut00}) visually inspected
 5$\times$5 degrees around the Galactic center and found 58 
star cluster candidates, Dutra \& Bica (\cite{dut01}) searched for 
embedded clusters and stellar groups in the area of known nebulae 
and added another 42 candidates. Bica et al. (\cite{bic03b}) 
summarized the new clusters and groups into a catalog containing 276 
objects. Dutra et al. (\cite{dut03}) searched the Southern Milky Way 
and added 179 more star clusters and groups. 
Automated searches based on stellar density contrast were carried
out by Carpenter et al. (\cite{car00}), Ivanov et al. (\cite{iva02};
hereafter Paper {\sc I}), Reyle \& Robin (\cite{rey02}), and Borissova 
et al. (\cite{bor03}; hereafter Paper {\sc II}), and yielded 22 new 
cluster candidates. 

The total number of new infrared-detected star clusters and stellar 
group candidates currently is about 500. Yet, this valuable dataset remains 
relatively unmined. The first step in the analysis is to identify true
clusters and then to determine their physical properties. This paper 
is part of a 
project to derive extinctions, distances, ages and total 
masses of these cluster candidates. In particular, we are addressing the question of whether the 
Milky Way is still forming massive clusters similar to the 
progenitors of present-day globular clusters. The best known examples of such 
objects are the Arches  (Nagata et al. \cite{nag93}; see Figer et al.
\cite{fig02} for latest review) and the Quintuplet clusters 
(Glass et al. \cite{gla87}; see also Figer et al. \cite{fig99b}). 
Are there more ``Arches'' in the Galaxy?

Here we report deep $J$, $H$, and $K_S$ imaging of 21 cluster
candidates, selected from Bica et al. (\cite{bic03a}) and Dutra et al. 
(\cite{dut03}). The majority of them are located in a 10$\times$10 
degree region around the Galactic Center. The targets were selected 
based on their classification in these papers, and on their 
resemblance to the Arches and the Quintuplet on the 2MASS images. 

The next section describes the data and the third section discusses the
confirmed clusters in detail. The fourth section includes comments 
on the spuriously identified clusters and objects whose nature cannot
be determined based on the currently available dataset. The last 
section is a summary of the results.

\section{Observations and data reduction}

Infrared imaging 
observations were carried out on July 31, 2003, Aug 1-2, 2003, and 
June 25, 2004, with the PANIC (Persson's Auxiliary Nasmyth Infrared Camera) 
near-infrared imager on the 6.5-meter 
Baade telescope at the Las Campanas Observatory. The instrument uses a
1024$\times$1024 HgCdTe Hawaii detector array. The scale is 0.125 
arcsec $\rm pixel^{-1}$, giving a total field of view of 
2.1$\times$2.1 arcmin. The observing log is given in 
Table~\ref{TableLog}. 

\begin{table*}
\begin{center}
\caption{Parameters of the cluster candidates and the log of 
observations.}
\label{TableLog}
\begin{tabular}{lccrcrrc}
\hline
\multicolumn{1}{c}{ID} &
\multicolumn{1}{c}{R.A.} &
\multicolumn{1}{c}{Dec.} {\hspace{5pt}}&
\multicolumn{1}{c}{{\it l}}{\hspace{5pt}} &
\multicolumn{1}{c}{{\it b}}{\hspace{5pt}} &
\multicolumn{1}{c}{Filter}&
\multicolumn{1}{c}{Exposure}&
\multicolumn{1}{c}{Date}\\
\multicolumn{1}{c}{} &
\multicolumn{2}{c}{(J2000.0)} &
\multicolumn{1}{c}{} &
\multicolumn{1}{c}{} &
\multicolumn{1}{c}{}& 
\multicolumn{1}{c}{$\sec$}&
\multicolumn{1}{c}{Observ.}\\
\hline 
$[$DB2000$]$\,7     & 17 50 06.1& $-$28 53 13 & 0.54& $-$0.81    & $K_S$ &  60&           01.08.2003\\
$[$DB2000$]$\,8     & 17 50 04.7& $-$28 52 40 & 0.55& $-$0.80    & $K_S$ &  60&           01.08.2003 \\
$[$DB2000$]$\,26    & 17 48 41.5& $-$28 01 42 & 1.12& $-$0.10    & $J$, $K_S$ & 600, 600& 01.08.2003 \\
$[$DB2000$]$\,40    & 17 53 15.9& $-$26 46 52 & 2.71& $-$0.34    & $K_S$ &  60&           01.08.2003\\
$[$DB2000$]$\,41    & 17 53 02.8& $-$26 39 26 & 2.84& $+$0.39    & $K_S$ &  60&           01.08.2003 \\
$[$DB2000$]$\,52    & 17 42 28.1& $-$29 56 23 & 358.78& $+$0.05  & $J$, $K_S$ & 300, 600& 31.07.2003\\ 
$[$DB2000$]$\,56    & 17 46 24.2& $-$29 22 19 & 359.71& $-$0.37  & $K_S$ &  50&           02.08.2003\\
$[$DB2000$]$\,58    & 17 45 00.1& $-$28 51 37 & 359.99& $+$0.15  & $K_S$ &  75&           02.08.2003\\
$[$DB2001$]$\,40    & 17 30 28.2& $-$34 41 30 & 353.42& $-$0.36  & $J$, $K_S$ & 300, 600& 01.08.2003\\
$[$DBSB2003$]$\,83  & 13 11 14.0& $-$62 45 04 & 305.20& $+$0.03  & $K_S$ &  30&           02.08.2003\\
$[$DBSB2003$]$\,84  & 13 11 47.0& $-$62 45 46 & 305.20& $+$0.02  & $K_S$ &  30&           02.08.2003\\
$[$DBSB2003$]$\,170 & 16 28 58.0& $-$49 36 27 & 334.72& $-$0.65  & $K_S$ & 300&           25.06.2004\\  
$[$DBSB2003$]$\,174 & 16 48 10.0& $-$45 21 29 & 340.05& $-$0.23  & $H$ & 300&             25.06.2004 \\  
$[$DBSB2003$]$\,177 & 17 04 13.0& $-$42 20 02 & 344.22& $-$0.60  & $H$, $K_S$ & 300, 300& 25.06.2004\\  
$[$DBSB2003$]$\,179 & 17 11 32.0& $-$39 10 38 & 347.58& $+$0.11  & $J$, $H$,$K_S$ & 300, 300, 300& 02.08.2003\\
$[$BDSB2003$]$\,9   & 18 17 53.0& $-$11 44 26 & 18.67 & $+$1.97  & $K_S$ & 150&           02.08.2003\\
$[$BDSB2003$]$\,101 & 17 25 34.0& $-$34 23 08 & 355.46& $-$0.38  & $K_S$ & 300&           01.08.2003\\
$[$BDSB2003$]$\,103 & 17 50 47.0& $-$31 16 34 & 358.57& $-$2.16  & $J$, $K_S$ & 150, 150& 02.08.2003\\
$[$BDSB2003$]$\,105 & 18 07 30.0& $-$25 44 30 & 5.20  & $-$2.60  & $K_S$ & 300&           25.06.2004\\  
$[$BDSB2003$]$\,106 & 18 01 35.0& $-$24 50 06 & 5.34  & $-$0.99  & $H$, $K_S$ & 300, 300& 25.06.2004\\  
$[$BDSB2003$]$\,107 & 18 00 42.0& $-$24 04 23 & 5.90  & $-$0.43  & $H$, $K_S$ & 300, 300& 25.06.2004\\  
\hline
\end{tabular}
\end{center}
\end{table*}

The observing strategy was typical for near-infrared imaging: we 
alternated between the object and a nearby sky, accumulating 1 to 3
minutes of integration at each pointing. At both the ``object'' and 
the ``sky'' pointing we jittered within $\sim$ 20-30 arcsec to 
minimize the effect of cosmetic defects and cosmic rays. 
For each target and filter the ``sky'' images were median-combined 
to create a ``super-sky'' image which was subtracted from the 
``object'' images. The ``super-skies'' were also used to create 
flats. Next, we shifted the ``object'' images to a common position, 
and combined them.

The stellar photometry of these sky-subtracted and combined images 
was carried out using ALLSTAR in DAOPHOT\,II (Stetson \cite{ste93}). 
We considered only stars with DAOPHOT errors less than 0.2 mag. 
The median averaged internal photometric errors are $0.03\pm0.02$ 
for the $J$, 
$H$, $K_S$ magnitudes brighter than 17 mag and $0.07\pm0.04$ for the 
fainter stars. We also added in quadrature an additional 
observational uncertainty of $\sim$0.03 mag due to the sky 
background variations. Finally, we replaced the brightest stars 
(usually with $K_S$$<$12 mag), that were saturated in our photometry,
with the 2MASS measurements.

The weather conditions were nonphotometric (typical seeing 1-1.2 arcsec) 
during all our observing 
runs, forcing us to calibrate the data by comparing our instrumental 
magnitudes with the 2MASS magnitudes of 10-25 stars per image, 
depending on the field crowding and the band. The standard error values for 
the coefficients are less than 0.03
for the zeropoint and less than 0.02 for the color term. 
In summary, our conservative
estimate of the total external errors of our photometry is 0.04-0.05 mag.

\section{Confirmed clusters}

\subsection{[DB2000]\,26}

[DB2000]\,26 is located on the sky close to the star-forming region 
Sgr\,D.  An H\,{\sc II} region is clearly visible on our images. 
The $K_S$ versus $J-K_S$ color-magnitude diagram is shown in 
Fig.~\ref{fig01}. 
If we adopt a cluster limiting radius of $r=0.5$ arcmin
and plot all the stars within this area they have 
$J-K_S$$>$3 mag. To obtain an estimate of the fore- and background 
contamination we define a non-cluster region
in the southern part of the images with an area equal to that of the cluster.
In this region we find a single 
``field'' star that satisfies the color criterion adopted for the cluster. We 
then statistically clean the cluster CMD, removing from the 
cluster color-magnitude diagram as many stars as on the ``field'' 
color-magnitude diagram (in this case, one star). 

\begin{figure}
\caption{The $K_S$ versus $J-K_S$ color-magnitude diagram 
of \object{[DB2000]\,26}. All stars in our photometry list 
are shown with solid dots. 
The field stars are marked with pluses, and the probable
cluster members with diamonds (see the text).
The typical photometric errors for different magnitude bins are 
shown on the right.
The unreddened Main Sequence (Schmidt-Kaler \cite{sch82}) is 
shown with a solid line, and with dashed lines 
for different reddening values, corresponding to E($B-V$)=2.5, 
E($B-V$)=7.2, and E($B-V$)=12 mag. 
The reddening vectors for B0\,V and B5\,V stars are also shown, for 
A$_V$=23 mag. 
The $(m-M)_0$=15 mag is adopted from Churchwell at al. (1990).
}
\label{fig01}
\end{figure}

To verify this selection we overploted all stars with $J-K_S$$>$3 
mag on the $K_S$ band image (Fig.~\ref{fig02}). The overdensity 
around the cluster center is obvious. We estimate that the diameter 
of the cluster is approximately 1.5 arcmin 
 (shown in Fig.~\ref{fig02} with a
large circle). The number of cluster members remaining after the 
``field'' subtraction described above was 43 (marked in 
Fig.~\ref{fig01} with diamonds).

\begin{figure}
  \caption{The $K_S$-band image of \object{[DB2000]\,26}. 
  The field of view is 2.1$\times$2.1 arcmin. North is up, 
  and East is to the right. The stars with $J-K_S>$\,3 mag 
  are marked with diamonds. The large cross and the circle 
  indicate the adopted cluster center and boundary. 
  }
  \label{fig02}
\end{figure}

We estimate the extinction toward the cluster members from the 
$K_S$ versus $J-K_S$ color-magnitude diagram (Fig.~\ref{fig01}), 
and adopt a distance modulus of $(m-M)_0$=15 mag (D=10 Kpc) based on 
water maser and ammonia emissions (\cite{chu90}). Since such sources are
commonly associated with regions of recent star formation, we 
assumed that they originate in the vicinity of the cluster. We 
compared the cluster sequence with the theoretical Main Sequence 
from Schmidt-Kaler (\cite{sch82}) for E($B-V$)=2.5, E($B-V$)=7.2, 
and E($B-V$)=12 mag. Throughout this paper we adopt R$_V$=3.2. 

The cluster distance is estimated with the 10th brightest star method described in Dutra et al. 
(\cite{dut03b}; see Section~\ref{DiscSumm} for discussion on the 
accuracy of this technique). Assuming that [DB2000]\,26 is not a massive 
cluster, and that the 10th brightest star corresponds to 
a B0\,V star with intrinsic color $(J-K)_0$=$-$0.16 and 
absolute magnitude $M_k$=$-$3.17 mag (Schmidt-Kaler 
\cite{sch82}), we calculate $E(B-V)$=7.0 and 
$(m-M)_0$=14.7 mag which is good agreement with the distance
determined by Churchwell at al. (\cite{chu90}). This is 
somewhat further than the distance to the star-forming region 
Sgr\,D, placed by Blum \& Damineli (\cite{blu99}) close to the 
Galactic Center but on the near side, leaving the question 
about the physical association between the cluster and Sgr\,D 
open.

The cluster members appear to be young main sequence stars. 
We exclude the presence of red giants as in some open clusters
because the associated H{\sc ii} region requires the presence of 
ionizing photons, and red giants appear only in clusters older
than 0.8 Gyr (Bertelli et al. \cite{ber94}) at which time the 
hot stars at the upper main sequence have evolved away.

Most of these stars occupy the locus between A$_V$=18 and 29 
mag. The seven brightest stars are exception to this with 
E($B-V$)$\sim$12 mag or A$_V$$\sim$38 mag.
We adopt as a mean color 
excess of the cluster E($B-V$)=7.2 mag or A$_V$=23 mag. The 
field stars appear to suffer between A$_V$=7 and 12 mag of 
visual extinction, and for them we adopt E($B-V$)=2.8 mag or 
A$_V$=9 mag. 

We estimate the age of the cluster and field 
stars by comparing the observed and theoretical luminosity functions 
(LF), closely following the method described in Porras et al. 
(\cite{por00}), based on a evolutionary sequence of six 
$J$-band LFs (Strom et al. 1993) for 0.3, 0.7, 1, 3,7 and 10 
Myr. We use the distance and the reddening listed above. 
Note that the distance to the cluster was adopted as an 
average distance to the field stars only for the purpose of 
comparing the LFs. The Kolmogorov-Smirnov test favors ages of 
1 Myr for the cluster and 3 Myr for the field population with  
$65\%$ and $55\%$ respectively (not very high) 
probability. The model-predicted and the observed 
$J$-band LFs are presented in Fig.~\ref{fig03},
where both have been normalized to a peak of one.

\begin{figure}
\vspace{4mm}
\caption{The normalized luminosity function of the members 
of \object{[DB2000]\,26} (top), and the stars in the 
selected comparison field (bottom).
The thick lines are the observed luminosity functions.
The thin lines are the theoretical models by Strom et al. 
(\cite{sto93}) for 1 and 3 Myr, for the field and for the 
cluster, respectively. }
\label{fig03}
\end{figure}

Finally, we attempt to determine the initial mass function 
slope of the 
cluster following the technique used in Paper {\sc II}. 
We adopted the 1 Myr isochrone, and counted 
cluster stars between reddening lines originating from positions 
on the isochrone for different initial masses.  The photometry 
of the cluster members suffer severe incompleteness for stars with 
masses lower than 10 solar masses due to the strong extinction in 
the $J$-band, leaving us with two mass bins. Therefore, we can
only put a limit on the IMF slope if $\Gamma$$\leq$$-$0.8$\pm$0.2. 
The uncertainty here represents only the formal fitting error.
The minimal total cluster mass, comprised of the mass of suspected
cluster members, is 780 solar masses. An integration of the 
extrapolated power-law IMF fit down to 0.8 $M_{\odot}$ yields a 
total mass of 2900 $M_{\odot}$  which should be considered an upper 
limit.

\subsection{[DB2000]\,52}

The cluster candidate [DB2000]\,52 is associated with the 
H\,{\sc II} region [LPH96]\,358.797+0.058, the IR source 
IRAS\,17392-2954, four radio-sources and the young stellar 
object ISOGAL-P~J174228.0-295614. Schultheis et al. 
(\cite{sch03}) reported $HK$-band spectroscopy of 107 
sources in the region. They measured $J$=13.9, $K_S$=10.55,
and $A_V$=24.30 mag for ISOGAL-P~J174228.0-295614, in 
agreement with our photometry: $J$=13.8 and 
$K_S$=10.43 mag.  

The cluster is located in an extremely crowded field, and while
individual stellar spectra would be required to accurately determine
cluster membership, statistical decontamination can nonetheless provide
an estimate of the true cluster CMD and LF.  Our statistical
decontamination is performed as follows.
First, we adopt a cluster limiting radius of r=0.5 arcmin. 
Next, a field with only non-cluster stars is defined 
as a circular annulus around the cluster with an inner radius 
of $r=1.1$ arcmin, and the same area as [DB2000]\,52. The 
$(J-K_S,K_S)$ CMDs of the ``cluster + field'' and the 
``field'' were gridded as shown in Fig.~\ref{Fig04}, and the 
stars in each box in the two diagrams were counted. Then, 
from the stars in each box of the ``clusters + field'' CMD 
we randomly removed  the same number of stars as in the 
corresponding box of the ``field'' CMD.  As the reddening in the field
is primarily in the cluster, foreground objects will be unaffected
regardless of their spatial location.  Background objects are 
different and are assumed to contribute negligibly.

\begin{figure}
\caption{The $(J-K_S,K_S)$ color-magnitude diagram for 
[DB2000]\,52: ``cluster + field'' on the left and
``field'' on the right. The grid for the CMD 
decontamination is shown (see the text for details).} 
\label{Fig04}
\end{figure}

The $J-K_S$ versus $K_S$ color-magnitude diagram is shown 
in Fig.~\ref{db2000_52cmd} and all measured stars are plotted 
as solid dots. The ``decontaminated'' cluster members are 
marked with diamonds. They are also shown as a diamonds on 
the $K_S$ band image (Fig.~\ref{db2000_52chart}).

\begin{figure}
\caption{The $K_S$ versus $J-K_S$ color-magnitude diagram 
of \object{[DB2000]\,52}. All stars in our photometry list 
are shown with solid dots. The statistically decontaminated cluster 
members are marked with diamonds. 
The typical photometric errors for different magnitude bins 
are shown on the right.
The unreddened Main Sequence (Schmidt-Kaler \cite{sch82}) is 
shown with the solid line, and with dashed lines for different 
reddening values, corresponding to E($B-V$)=6, E($B-V$)=7.6 
and E($B-V$)=9 mag. 
The reddening vectors for B0\,V and B5\,V are also shown for 
A$_V$=23 mag. 
The adopted distance modulus is $(m-M)_0$=11.5 mag.
}
\label{db2000_52cmd}
\end{figure}

We estimate the cluster distance using the 
10th brightest star method (Dutra et al. \cite{dut03b}),
finding $E(B-V)$=8.7, $A_V$=27.8, and 
$(m-M)_0$=11.37 mag. Assuming the extinction value obtained 
for ISOGAL-P~J174228.0-295614 gives $(m-M)_0$=11.67 mag.

Thus, for this cluster we found 
reddening between $A_V$=19 and 29 mag and several stars with
infrared excess. The cluster stars are young main sequence 
stars. We adopted the average distance modulus 
$(m-M)_0$=11.5 mag or D$\sim$2 kpc. The heavy background and 
foreground contamination and the uncertain distance prevent 
us from estimating the age and 
the IMF slope of this cluster. The minimum total mass, 
comprised of suspected member stars, is 600-700 $M_{\odot}$ ,
similar to the previous object.

An upper limit to the cluster mass can be estimated by
adopting the Salpeter law, normalizing it to the upper three mass bins 
(encompassing stars with masses about 10 solar masses), 
and integrating down to 0.8 solar masses. This 
yields a total cluster mass of 5000 $M_{\odot}$ or 
less. This is a very conservative limit because about 20\%
 of this mass is below 1 solar mass, and the 
IMF is known to flatten in this range. It is impossible 
to derive the IMF  slope because only the two upper mass 
bins are complete in the $J$-band due to the extreme 
extinction; we therefore assume the Salpeter slope. 

\begin{figure}
  \caption{The $K_S$ band image of \object{[DB2000]\,52}. 
  The field of view is 2.1$\times$2.1 arcmin. North is up, 
  and East is to the right. The statistically decontaminated 
  cluster stars are marked with diamonds. The large cross 
  and the circle indicate the adopted cluster center and 
  boundary. 
  }
  \label{db2000_52chart}
\end{figure}

\subsection{[DB2001]\,40 or [BDB2003]\,G353.42-00.36G353.42-00.36}

[DB2001]\,40 appears to be associated with the H\,{\sc II} region 
[FC2000]\,G353.41$-$0.36. The Southern part of our images is 
free of nebulosity and we adopted it as a comparison 
field. The $K_S$ versus $J-K_S$ color-magnitude diagram is 
shown in Fig.~\ref{db2001_40cmd}, and all measured stars are 
plotted as solid dots. The stars from our ``comparison'' 
field are plotted with pluses. As can be seen, most of them 
form a sequence with $J-K_S$$<$3 mag.

\begin{figure}
\caption{The $K_S$ versus $J-K_S$ color-magnitude diagram 
of \object{[DB2001]\,40}. All stars in our photometry list 
are shown with solid dots. The statistically decontaminated cluster 
members are marked with diamonds,
and field stars are marked with +'s. 
The typical photometric errors for different magnitude bins 
are shown on the right.
The unreddened Main Sequence (Schmidt-Kaler \cite{sch82}) is 
shown with the solid line, and with dashed lines for different 
reddening values, corresponding to E($B-V$)=2.7, E($B-V$)=7.2 
and E($B-V$)=9.5 mag. 
The reddening vectors for B0\,V and B5\,V stars are also shown, for 
A$_V$=24 mag. 
The adopted distance modulus is $(m-M)_0$=15.7 mag.
}
\label{db2001_40cmd}
\end{figure}

We marked all stars which have $J-K_S >$\,3 mag on the 
$K_S$-band image (Fig.~\ref{db2001_40chart}). An overdensity of stars 
around the cluster center is obvious. We estimated the cluster 
diameter as $\sim$0.6-0.7 arcmin (shown in Fig.~\ref{db2001_40chart} with 
a large circle). There are 13 suspected cluster members left 
after the CMD decontamination. They are marked with diamonds 
in Fig.~\ref{db2001_40cmd}.

\begin{figure}
  \caption{The $K_S$ band image of \object{[DB2001]\,40}. 
  The field of view is 2.1$\times$2.1 arcmin. North is down, 
  and East is to the left. The stars with $J-K_S$$>$3 mag
  are marked with diamonds. The large cross and the circle 
  indicate the adopted cluster center and boundary. 
  }
  \label{db2001_40chart}
\end{figure}

Here again we are forced to apply the 10th brightest star 
method to obtain an approximate distance estimate. It appears
that [DB2001]\,40 is not a massive cluster, and following 
Dutra et al. (\cite{dut03b}) we assume that the 10th 
brightest star is a B0\,V star. From the difference between 
the apparent and the intrinsic color and the apparent and the
intrinsic magnitude we derive $E(B-V)$=7.3, $A_V$=23.4 and 
$(m-M)_0$=15.7 mag  (D=13.8 kpc). This distance is in 
good agreement with that from  Walsh et al. (1997), who
obtained a kinematical distance of 15.6 kpc for IRAS 17271-3439.
In their study of ultracompact (UC) HII regions they derived the 
distances kinematically, using the 6.669\,GHz methanol 
maser emission velocity.
The small number of cluster members renders any IMF 
slope derivation meaningless. The observed members comprise a total mass of 
about 400 solar masses.
The upper mass limit of this cluster is 450 solar 
masses but it is more uncertain in comparison with the
other confirmed clusters because of the small number of
potential cluster members.

\subsection{[DBSB2003]\,177}

Walsh et al. (1997) found an ultra-compact H\,{\sc II} region 
in the vicinity of [DBSB2003]\,177.
The region is associated with a red star ($J-K_S$=4.7). They detected a 
methanol and OH maser source 5 arcsec to the North, 
with no apparent near-infrared counterpart. 

To decontaminate 
the cluster's CMD we followed the same method as for 
[DB2000]\,52, adopting a cluster limiting radius of $r$=0.5 
arcmin. The ``field'' population was determined from a circular 
annulus around the cluster, with an inner radius $r$=1.1 arcmin 
and the same area as the cluster. The gridded $(H-K_S,K_S)$ 
CMDs of ``cluster + field'' and ``field'' are shown in 
Fig.~\ref{db177_field}.

\begin{figure}
\caption{The $(H-K_S,K_S)$ color-magnitude diagram for 
[DBSB2003]\,177: ``cluster + field'' on the left 
and ``field'' on the right. The grid for the CMD 
decontamination is shown (see the text for details).} 
\label{db177_field}
\end{figure}

The $K_S$ versus $H-K_S$ color-magnitude diagram of all 
stars in the images is shown in Fig.~\ref{db177_cmd}. The 23 
suspected members remaining after the ``decontamination'' 
are marked with diamonds. Their spatial location (again,
marked with diamonds) is shown on the $K_S$ band image in 
Fig.~\ref{db177_chart}.

\begin{figure}
\caption{The $K_S$ versus $H-K_S$ color-magnitude diagram 
of [DBSB2003]\,177. All stars in our photometry list 
are shown with solid dots. The decontaminated cluster 
members are marked with diamonds. The unreddened Main
Sequence (Schmidt-Kaler \cite{sch82}) is 
shown with solid line, and with dashed lines for different 
reddening values, corresponding to E($B-V$)=1.5, 
and E($B-V$)=5 mag. 
The reddening vectors for B0\,V and B5\,V stars are also shown, for 
A$_V$=19 mag. 
The adopted distance modulus is $(m-M)_0$=17 mag.
}
\label{db177_cmd}
\end{figure}

Again, we use the 10$^{th}$ brightest star method to estimate 
the cluster distance, under the assumption that [DBSB2003]\,177 
is not a massive cluster  and the 10th brightest star 
corresponds to a B0\,V star with $(H-K)_0$=$-$0.04 and 
$M_K$=$-$3.17 mag (Schmidt-Kaler \cite{sch82}). We obtain
$E(B-V)$=5, $A_V$=16 and $(m-M)_0$=16.3 mag (18 kpc). 
This distance is close to the kinematical distance of 16.6 
kpc towards IRAS\,17006-4215, located in the vicinity of
the cluster (Walsh et al. \cite{wal97}). Assuming an age of 
1 Myr, the suspected cluster members comprise 200-400 $M_{\odot}$. 
As before, we refrain from further analysis because 
of the uncertain distance. 
The upper mass limit for this cluster is 1200 $M_{\odot}$,
based on an integration of the LF down to 0.8 $M_\odot$, assuming a Salpeter slope.

\begin{figure}
  \caption{The $K_S$ band image of \object{[DBSB2003]\,177}. 
  The field of view is 2.1$\times$2.1 arcmin. North is up, 
  and East is to the right. The statistically decontaminated 
  cluster stars are marked with diamonds. The large cross 
  and the circle indicate the adopted cluster center and 
  boundary. 
  }
  \label{db177_chart}
\end{figure}

\subsection{[DBSB2003]\,179}

A $K_S$-band image of the candidate is shown in 
Fig.~\ref{fig12}. The overdensity of stars is obvious.

\begin{figure}
  \caption{The $K_S$ band image of \object{[DBSB2003]\,179}. 
  The field of view is 2.1$\times$2.1 arcmin. North is down, 
  and East is to the left. The large cross 
  and the circle indicate the adopted cluster center and 
  boundary. 
  }
  \label{fig12}
\end{figure}

The decontamination procedure described previously left 152 
potential cluster members within 0.5 arcmin of the cluster 
center. The $K_S$ versus $J-K_S$ color-magnitude diagram is 
plotted in Fig.~\ref{fig13}.

\begin{figure}
\caption{The $K_S$ versus $H-K_S$ color-magnitude diagram 
of [DBSB2003]\,179. All stars in our photometry list 
are shown with solid dots. The statistically decontaminated cluster 
members are marked with diamonds, field stars are marked with +'s, and open circles are
for bright stars added from 2MASS photometry.
The unreddened Main Sequence (Schmidt-Kaler \cite{sch82}) is 
shown with the solid line, and with dashed lines for different 
reddening values, corresponding to E($B-V$)=2.7, 
and E($B-V$)=6 mag. 
The reddening vectors for B0\,V and B5\,V stars are also shown, for 
A$_V$=20 mag. 
The adopted distance modulus is $(m-M)_0$=13.5 mag.
}
\label{fig13}
\end{figure}

The 10th brightest star method yields $E(B-V)$=6, $A_V$=19 
and $(m-M)_0$=13.5 mag (5 kpc).
With such obtained parameters, a comparison with the theoretical luminosity 
functions of Strom et al. (1993) gives an age of 
the cluster 7-8 Myr. Predicted and observed
$J$ band LFs are presented in Fig.~\ref{fig14}. 

\begin{figure}
\vspace{4mm}
\caption{The normalized luminosity function of the [DBSB2003]\,179 cluster
(thick line), and the theoretical 7 Myr model by Strom et. al (\cite{sto93}) (thin line).}
\label{fig14}
\end{figure}

Given these parameters, the 
total mass of the cluster members is 
1100-1900 $M_{\odot}$.
The upper mass limit for this cluster is 5500 $M_{\odot}$.

\subsection{[BDSB2003]\,106}

The $K_S$-band image of [BDSB2003]\,106 is shown in 
Fig.~\ref{fig15}. There is an obvious stellar overdensity in
the center of the field, associated with extended emission
from a gaseous nebulosity.
  
\begin{figure}
  \caption{The $K_S$ band image of \object{[DBSB2003]\,106}. 
  The field of view is 2.1$\times$2.1 arcmin. North is up, 
  and East is to the right. The large cross 
  and the circle indicate the adopted cluster center and 
  boundary. 
  }
  \label{fig15}
\end{figure}

We followed the same analysis procedure as for the previous 
objects, and after the decontamination we selected 152 
possible cluster members within 0.5 arcmin from the cluster
center. The $K_S$ versus $H-K_S$ color-magnitude diagram is 
shown in Fig.~\ref{fig16}.

\begin{figure}
\caption{The $K_S$ versus $H-K_S$ color-magnitude diagram 
of [DBSB2003]\,106. All stars in our photometry list 
are shown with solid dots. The statistically decontaminated cluster 
members are marked with diamonds. 
The unreddened Main Sequence (Schmidt-Kaler \cite{sch82}) is 
shown with the solid line, and with dashed lines for different 
reddening values, corresponding to E($B-V$)=1.5, 
and E($B-V$)=6.5 mag. 
The reddening vectors for B0\,V and B5\,V stars are also shown, for 
A$_V$=20 mag. 
The adopted distance modulus is $(m-M)_0$=15.2 mag.
}
\label{fig16}
\end{figure}

The 10th brightest star method gives $E(B-V)$=6.5, 
$A_V$=20.8, and $(m-M)_0$=15.2 mag (11 kpc). 
Again, we refrained from further analysis but the presence 
of the emission nebula around the cluster indicates that 
it is 5 Myr of age or younger. The mass of 
the detected cluster members is 600-800 solar masses.
The upper mass limit for this cluster is 4000 solar masses.

\subsection{[BDSB2003]\,107}

The $K_S$-band image of [BDSB2003]\,107 is shown in 
Fig.~\ref{fig17}. This is the largest cluster candidate in our 
sample, and here we adopted a radius of 1 arcmin. The object
is probably associated with the H\,{\sc II} region 
WC89\,005.09$-$0.39A, so the object must be younger than 5 Myr.

\begin{figure}
  \caption{The $K_S$ band image of \object{[DBSB2003]\,107}. 
  The field of view is 2.1$\times$2.1 arcmin. North is up, 
  and East is to the right.  The large cross 
  and the circle indicate the adopted cluster center and 
  boundary. 
  }
  \label{fig17}
\end{figure}

The CMD decontamination leaves 309 possible cluster 
members. The $K_S$ versus $H-K_S$ color-magnitude diagram is 
shown in Fig.~\ref{fig18}. The 10th brightest star method 
gives $E(B-V)$=5.7, $A_V$=18.2, and $(m-M)_0$=13.8 mag 
(5.8 kpc). The total mass of the measured cluster members 
is 800-1000 solar masses. The upper mass limit for this cluster 
is 4500 solar masses.

\begin{figure}
\caption{The $K_S$ versus $H-K_S$ color-magnitude diagram 
of [DBSB2003]\,107. All stars in our photometry list 
are shown with solid dots. The decontaminated cluster 
members are marked with diamonds. 
The unreddened Main Sequence (Schmidt-Kaler \cite{sch82}) is 
shown with solid line, and with dashed lines for different 
reddening values, corresponding to E($B-V$)=1.5, 
and E($B-V$)=5.7 mag. 
The reddening vectors for B0\,V and B5\,V are also shown, for 
A$_V$=20 mag. 
The adopted distance modulus is $(m-M)_0$=13.8 mag.
}
\label{fig18}
\end{figure}

\section{Spurious candidates and uncertain detections}

Our data reveal that some candidates are actually not 
clusters. Most often they turned out to be just one or 
more bright stars, well above the 2MASS saturation limit.
Located close enough, such stars closely resemble compact
clusters at the 2MASS resolution. We consider groups of up to a dozen bright 
stars not to be clusters unless a concentration of some 
fainter stars is also observable. In this section we also list 
objects for which our data does not allow us to derive a 
definitive conclusion. A mosaic of our cluster images is shown 
in Fig.~\ref{fig33}. In addition, one of the cluster 
candidates appeared to be a well-known globular cluster.

\begin{figure}
  \caption{Mosaic $K_S$-band images of the spurious 
  clusters and objects with uncertain nature.
  The images are 2.1 arcmin on the side.
  }
  \label{fig33}
\end{figure}

1. [DB2000]\,7 and [DB2000]\,8: 
On our 160-second $K_S$-band image the cluster candidate 
[DB2000]\,7 is resolved into 7-8 bright stars. There is no 
overdensity of fainter stars close to the center of the object 
in comparison with the periphery of the image. There are no 
associated H\,{\sc II} regions or IRAS sources, unlike the 
typical young clusters. The same field contains the cluster 
candidate [DB2000]\,8, - also resolved into several bright 
stars, but showing no overdensity of fainter stars. The 
cluster-like appearance of [DB2000]\,7 and [DB2000]\,8 on the 
2MASS images is likely due to the limited angular resolution.

2. [DB2000]\,40:
Based on one 60-second $K_S$-band image, the cluster candidate 
[DB2000]\,40 appears to be a group of 6-7 bright stars. There 
is no overdensity of fainter stars, and no associated 
H\,{\sc II} regions or an IRAS source. 

3. [DB2000]\,41 was resolved into two bright stars on a 
60-second $K_S$-band image.

4. [DB2000]\,56: 
Dutra et al. (\cite{dut03}) classifies this object as  deeply 
embedded in dust and gas. Our data (50-second $K_S$-band image) 
shows that this is one bright star in the ultacompact 
H\,{\sc II} region. 

5. [DB2000]\,58:
Dutra et al. (2003) report that this cluster is related to the 
optical H\,{\sc II} region Sh\,2-17. Our 75-second $K_S$-band 
image indeed shows a conglomerate of about a dozen bright stars, 
but no overdensity of fainter stars.

6. [DBSB2003]\,83 and [DB2000]\,84: both objects were resolved 
into single bright stars, embedded in an ultracompact H\,{\sc II} 
region.

7. [DBSB2003]\,170 is an ultracompact H\,{\sc II} region, which 
contains several bright stars. No concentration of faint stars
is present.

8. [DBSB2003]\,174 is a group of bright stars associated 
with an H\,{\sc II} region, but there is no overdensity of faint 
stars.

9. [BDSB2003]\,9 shows a concentration of stars, associated with 
a H\,{\sc II} region. The object has the appearance of a star 
forming region but we have only a single 150-second $K_S$-band 
image, preventing us from  carrying out 
further analysis. This object needs
further observations.

10. [BDSB2003]\,101 is resolved into two bright stars on our 
300-second $K_S$-band image. 
 
11. [BDSB2003]\,103 is the well-known globular cluster Terzan\,6.

12. [BDSB2003]\,105 is a single bright star.

\section{Uncertainties in the distance estimates\label{DiscSumm}}

The 10th brightest star method described in Dutra et al. 
(\cite{dut03b}) was used to determine the cluster distance. 
This method is meant to improve  statistics when estimating the 
distance of a cluster.  Rather than just 
looking at the magnitude of the brightest star, which is very uncertain due 
to the small number of stars, they look farther down the MS where there 
are more stars, and hence the difference between the nth and (n+1)th star 
is much smaller.  One would like to look at the entire population, or at 
least the entire upper MS as a whole to get the best results.  
This technique gives only an approximate estimate, 
with systematic and random errors.

The former ones can be estimated internally by comparison 
between the distances derived from the assumptions that the 
cluster is massive versus non-massive, which corresponds to the 
10th star being O5\,V or B0\,V, respectively. This leads to a 
difference in the distance modulus of $\sim$1.5 mag, or factor 
of $\sim$4. Externally, the uncertainty can be estimated from 
comparison with distances obtained from other techniques, which 
are not available for the majority of the objects.

The random errors can be quantified by assuming Poisson 
statistics in the number of stars. As was mentioned above, 
the method assumes that the 10th brightest stars is a B0\,V. 
Re-stating this, there are 10$\pm$3 stars brighter than the 
B0\,V star in the cluster. Therefore, we assumed that the 
1$\sigma$ random error is the difference between the 
extinction and distance values derived assigning the B0\,V 
spectral type to the 10th brightest star, and to the 13th 
star. In effect, here we simply propagated the random sampling 
errors. Of course, this is only an approximate estimate, due 
to the small number statistics. The typical uncertainty in the
distance modulus due to the statistical errors is about 0.5 
mag or a factor of $\sim$1.5.

There is an error component associated with the differential 
reddening within the cluster. This error can be estimated from 
the width of the ``decontaminated'' cluster sequences on the 
color-magnitude diagrams. Discarding the outliers, which are 
most likely stars with near-infrared excess, the typical 
half-width of the cluster sequences is $\Delta(J-K_S)$$\sim$0.5 
mag, corresponding to A$_K$$<$0.5 mag. This is comparable to the 
uncertainties of our distance modulus due to the random errors 
discussed above.

Summarizing all uncertainties discussed above, we obtain 
$\sigma$\{(m-M)$_0$\}$\sim$2.5 mag or a factor of $\sim$10. 
However, these calculations are made under the assumption that 
the 10th brightest 
star is indeed on the main sequence, 
and if it is not, this could introduce much larger errors into the 
distance estimate.
Therefore, the distances derived from this method have to be 
treated with extreme caution.

\section{Discussion and summary}

We have presented deep near-infrared imaging with angular resolution 
superior to 2MASS which indicates that the identification of
clusters on the 2MASS images alone is far from reliable. The 
2MASS images suffer from large pixel size and poor seeing conditions
that have made many bright single or double stars appear as
compact clusters. Furthermore, there are a number of cluster 
candidates that consist of a few - usually fewer than a dozen 
bright stars, - that are not associated with concentrations of
fainter stars. Here we refrain from classifying those as 
clusters. However, they may be remnants of clusters, i.e. 
groups of common proper motion that have been ejected from 
larger stellar clusters. There is also the possibility that they
may be genuine ``clusters'' with extremely peculiar IMFs with 
high lower cut-off. 

Seven of the 21 observed objects appear to be 
typical young clusters. Most 
of them show extended gas emission, suggesting the presence of 
OB stars. Whenever possible, we estimated the mass of the 
clusters or at least of the probable cluster members for 
which we have photometry. These estimates range from a few 
hundred to a few thousand solar masses. None of the objects 
appear to be comparable in appearance and total mass to the 
Arches -- estimated to have a dynamical mass of 
7$\times$10$^4$\,$M_\odot$ (Figer et al. \cite{fig99a}) -- or 
the Quintuplet clusters -- estimated $\sim$10$^4$\,$M_\odot$ 
from an integration of the cluster IMF (Figer et al. \cite{fig99b}).

A special word of caution is necessary when discussing the 
masses of the clusters. The photometry alone is not sufficient
to derive the distances to the clusters, and we were forced
to apply a technique based on assuming a spectral type for the 
10th brightest cluster member. There is a number of 
uncertainties in this method, due to the age, foreground and 
background contamination, and even the variations in the 
cluster population due to small number statistics. 
Spectroscopic observations to derive the spectral types and 
distances would be much more accurate, but this is beyond the 
scope of this paper.

\begin{acknowledgements}
This publication makes use of data products from the Two 
Micron All Sky Survey, which is a joint project of the 
University of Massachusetts and the Infrared Processing 
and Analysis Center/California Institute of Technology, 
funded by the National Aeronautics and Space Administration 
and the National Science Foundation. This research has made 
use of the SIMBAD database, operated at CDS, Strasbourg, 
France. The authors gratefully acknowledge the comments 
by an anonymous referee. 

J.B., D.M. and D.G. are supported by FONDAP Center for Astrophysics grant 
number 15010003. 
\end{acknowledgements}

\end{document}